\documentclass[runningheads]{llncs}
\usepackage[T1]{fontenc}
\usepackage{graphicx}

\usepackage{amsmath}
\usepackage{array}
\usepackage[labelfont=bf, textfont=normalfont]{caption}
\usepackage{float}
\usepackage{longtable}
\usepackage{multirow}
\usepackage{tabularx}
\usepackage{url}	    

\usepackage[hidelinks]{hyperref}
\usepackage{xurl}
\usepackage{makecell}

\begin{document}
\title{Evaluating Calibration-Based Digital Twins for IBM Quantum Hardware Simulation}
\titlerunning{Evaluating Quantum Calibration-Based Digital Twins}
%
\author{Edgars~Bautra\orcidID{0009-0002-3632-5897} \and
Maksims~Dimitrijevs\orcidID{0000-0002-4225-7889} \and
Abuzer Yakary{\i}lmaz\orcidID{0000-0002-2372-252X}}
\authorrunning{Bautra, Dimitrijevs, Yakary{\i}lmaz}
%
\institute{Center for Quantum Computer Science, University of Latvia, Latvia
\email{edgars.bautra12@gmail.com, \\ \{maksims.dimitrijevs,abuzer.yakaryilmaz\}@lu.lv}\\
}
\maketitle              
\begin{abstract}
We evaluate calibration-based digital twins for IBM Quantum hardware, aiming to reproduce hardware measurement outcomes on classical simulators. We present a workflow that builds twins from downloadable calibration CSV files by mapping coherence times, gate and readout error rates, and operation durations to thermal-relaxation, depolarizing, and readout error channels, while reconstructing a directed coupling map to restore connectivity constraints during transpilation. We compare four twin variants (CSV-built, backend-derived simulator, backend-derived noise model, and fake-backend snapshots) under a common execution and validation protocol. Experiments on two IBM QPUs, \textit{ibm\_brisbane} and \textit{ibm\_sherbrooke}, use randomized five-qubit circuits of depths 10, 20, and 30 across four optimization levels. Weighted Jaccard similarity indicates that twins constructed from downloadable calibration CSV data often achieved the closest agreement with hardware, while backend-derived twins provided competitive and practical baselines. The results further show that agreement depends on both the target device and the transpilation settings, underscoring the need to validate digital twins for the specific execution setup rather than assuming transferability across devices.  

\keywords{Quantum computers \and digital twin \and noisy simulation \and  Qiskit }
\end{abstract}

\section{Introduction}

In today's world, quantum computers as a technology are undergoing rapid development. Although there are large companies, such as \textit{IBM}, that offer access to these computers through their cloud services \cite{ref_nguyen_qcloud}, demand still exceeds availability, leading to long queues of submitted circuits, because of which certain restrictions have been introduced, such as limited available computational time \cite{ref_romero_noisevalidation}.

Quantum computer simulators can be of considerable help with this situation, allowing anyone to create and run circuits on everyday devices -- classical computers, based on recreated properties of quantum physics \cite{ref_cicero_simulation}. In addition, less computational resources of quantum computers are being used by reducing the workload from unnecessary jobs that can be completed on simulators. Even so, there is one characteristic property of quantum computers that can cause the expected results to change in unpredictable ways -- noise, including decoherence, gate errors, and readout errors \cite{ref_perez_ibmreliability}, which makes ideal simulation insufficient. This creates a need for calibration-based digital twins of quantum computers, i.e., simulator configurations that use noise models built from device-specific calibration information to approximate the noisy behavior of a corresponding real device. Availability of digital twins can compensate the restricted availability to real Quantum Processing Units (QPUs), as well as allow one to do preliminary testing before running jobs on QPUs, thus sparing quantum computational resources.

Recent research explicitly frames calibration-based emulation as a digital twin problem. In particular, calibration-driven emulation of superconducting transmon devices has been presented as a way to build a digital twin from calibration data and additional benchmarking information, showing that hardware-aware emulation can improve agreement with the behavior of a real device \cite{ref_mueller_digitaltwin}. At the same time, Qiskit already provides practical support for noisy simulation based on backends, and IBM makes QPU calibration data downloadable as CSV \cite{ref_ibm_noisemodels,ref_ibm_backenddetails}. It is important to note that Qiskit can automatically configure a simulator to mimic an available backend via its specific methods such as \texttt{AerSimulator.from\_backend()}, while constructing a twin directly from downloadable calibration CSV files is not provided as a native end-to-end workflow in Qiskit. CSV files open up the possibilities to work with digital twins based on a wider range of QPUs, including ones not available to user.

Calibration CSV data derived from IBM has been used in quantum digital twin and simulation contexts. For example, Luo et al. imported calibration data of IBM QPUs collected as CSV files, and extend the twin concept to quantum-cloud operations like job lifecycle and fidelity estimation \cite{ref_luo_digitaltwinclouds}. Simulation frameworks such as ``iQuantum'' support importing IBM-style calibration CSV datasets for modeling quantum computing environments \cite{ref_cloudslab_iquantum}. These studies show the value of calibration-aware simulation; however, to the best of our knowledge, they do not describe a practical, Qiskit-native workflow for building circuit-executable twins directly from downloadable IBM calibration CSV files and validating them against IBM hardware measurement distributions. This leaves a gap for users seeking simulators that capture both device noise and execution constraints.

We address this gap by developing and validating calibration-based digital twins of IBM Quantum hardware using Qiskit and downloadable IBM calibration CSV data. We build noise models from calibration attributes and reconstruct the directional coupling information needed for device-faithful simulation. We compare this CSV-based approach with backend-derived and snapshot-based alternatives and validate all twin variants against real IBM QPU outputs using Weighted Jaccard Similarity. The results show that twins based on CSV are feasible and often highly competitive, while backend-derived twins remain strong practical baselines.

The code, results of our experiments, and device calibration data files are publicly available at \url{https://github.com/BautraE/precise-qpu-simulation-extras}.

The remainder of the paper is organized as follows. Section~2 presents the main contributions of the paper. Section~3 describes the technical details of the conducted research. Section~4 presents the results and discussion, including implications, limitations, and concluding remarks.

\section{Main Contributions}

The main contributions of this paper are as follows:

\begin{itemize}
    \item We present a practical method for constructing a QPU digital twin from downloadable IBM calibration CSV data, including the use of calibration attributes for noise-model creation and a workaround for reconstructing directional coupling information required for device-faithful transpilation and simulation.

    \item We formalize and compare four digital twin variants based on different sources of device and noise information -- custom noise models built from downloadable calibration CSV files, backend-derived simulator and noise-model variants, and fake-backend snapshots.

    \item We introduce a validation procedure based on Weighted Jaccard Similarity for comparing measured state-count outputs from QPU runs and simulator runs under a common evaluation protocol.

    \item We investigate the impact of shot count on result stability and select a practical shot budget for calibration-based digital twin evaluation.

    \item We provide an empirical validation study on two IBM Quantum QPUs, using three randomly generated five-qubit circuits and four transpiler optimization levels to compare hardware results with simulator results across all twin variants. 
\end{itemize}

\section{Technical Details}

\subsection{Problem setting and digital twin definition}

Local simulation allows us to compensate for constraints on access to real quantum hardware. When preparing and testing circuits before execution on a real QPU, the usefulness of local simulation depends on how closely the simulator reproduces the noisy behavior of the QPU.

In this paper, the problem is formulated as a \emph{digital twin validation} problem. We evaluate the output similarity between real-device runs and simulator runs produced by multiple twin variants (agreement of measured circuit-execution results under noisy conditions).

We define \emph{calibration-based digital twin} of a QPU as a simulator instance parameterized with device information in order to approximate the noisy execution behavior of that device. The digital twin consists of:
\begin{itemize}
    \item a simulator backend configuration;
    \item a noise model;
    \item device execution constraints (e.g., basis gates and coupling topology) relevant for transpilation and simulation;
    \item an execution workflow that returns measurement counts comparable to QPU runs.
\end{itemize}


\subsection{Validation Goal and Acceptance Criteria}

Our goal is to validate whether the twin is useful for reproducing circuit outputs observed for a QPU, i.e., whether the received results from a circuit execution are similar enough between a QPU and a simulator that is based on it. The comparison is performed at the level of measured state-count outputs, and we use a similarity metric defined in Section \ref{similarity_metric}.

We define the following similarity thresholds for digital twins:
\begin{itemize}
    \item 95\% as near-identical to QPU;
    \item 90\% as close match to QPU;
    \item 85\% as usefully similar to QPU.
\end{itemize}

95\% acceptance threshold is not uncommon, for example, in \cite{ref_bertomeu_maestro}, authors set acceptance threshold of $0.95$ for simulation fidelity. Thresholds of 90\% and 85\% have reasonable total variation distance ($\leq 5.26\%$ and $\leq 8.11\%$, respectively).

\subsection{Twin Variants and Execution Workflow}

This subsection summarizes the digital-twin variants compared in the study and the common workflow used to construct, execute, and validate them. We evaluate four twin variants within one common framework.This allows us to attribute the differences in output agreement to the twin-construction approach rather than to the changes in the comparison procedure. 

The evaluated digital twin variants are: 
\begin{itemize}
    \item Custom noise models from device calibration data (created from downloadable \textit{CSV} files); 
    \item Automatically obtained simulator instances that are based on a specific IBM Quantum QPU; 
    \item Automatically obtained noise model instances that are based on a specific IBM Quantum QPU. It is closely related to the simulator instance (previous point), but not identical (the simulator variant also carries backend execution constraints such as basis gates and coupling map);
    \item Automatically obtained fake provider simulator instances that are based on QPU snapshots -- \textit{FakeBrisbane} and \textit{FakeSherbrooke}. 
\end{itemize}

The workflow has four main stages:

\begin{enumerate}
    \item \textbf{Device-data acquisition and preparation.}  
    Device-related information is collected from the target QPUs (basis gates and calibration data in the form of CSV files).

    \item \textbf{Twin construction.}  
    Instantiating simulators involves both manually creating new simulators and obtaining already prepared ones by IBM. Created simulators are then configured with noise models that are also either created or simply obtained.

    \item \textbf{Circuit execution.}  
    The selected benchmark circuits are transpiled and executed on the real QPU and each twin variant.

    \item \textbf{Validation comparison.}  
    Similarity matrices and summary statistics for evaluation are produced based on measurement count outputs.
\end{enumerate}


\subsection{Calibration Data Extraction and Preprocessing}
\label{calibration_data_extraction_and_preprocessing}

Even though it is currently possible to simulate quantum computers and create noise models that mimic effects of decoherence on classical computers, specific attribute values must be used when defining certain types of noise in order for the simulator results to align with those of real QPUs. In regards to this, \textit{IBM} regularly conducts measurements and calculations to produce calibration data for every single qubit of their QPUs. During preprocessing, all time-valued attributes are converted to a common unit before being used in noise-parameter calculations. In particular, coherence times are reported in microseconds, while operation durations (readout and gate lengths) are reported in nanoseconds. The following attributes are available in the downloadable calibration data CSV files from the \textit{IBM Quantum} platform, and their names are taken directly from these files \cite{ref_ibm_backenddetails}: 
\begin{itemize}
    \item \textbf{Qubit identification and status attributes:}
    \begin{itemize}
        \item \textbf{Qubit} -- index of qubit to which the calibration values apply;
        \item \textbf{Operational} -- boolean value that shows if the qubit is currently usable.
    \end{itemize}
    \item \textbf{Coherence time attributes} are used for thermal relaxation noise:
    \begin{itemize}
        \item \textbf{T1 (us)} -- energy-relaxation time;
        \item \textbf{T2 (us)} -- phase-coherence time.
    \end{itemize}
    \item \textbf{Control attributes} are relevant to hardware control and pulse-level behavior, but they are not used in our current gate-level noise-model construction:
    \begin{itemize}
        \item \textbf{Frequency (GHz)} -- each qubit has its own frequency that needs to be used, when applying a certain gate;
        \item \textbf{Anharmonicity (GHz)} -- “the difference in energy between the first and second excited states of the qubit” \cite{ref_ibm_backenddetails}.
    \end{itemize}
    \item \textbf{Readout attributes:}
    \begin{itemize}
        \item \textbf{Readout assignment error} -- the probability to measure the wrong value for a qubit;
        \item \textbf{Prob meas0 prep1} (\textbf{Prob meas1 prep0}) -- probability that the measurement of a qubit's value returns '0' ('1') immediately after it was prepared as state \( \vert1\rangle \)  (\( \vert0\rangle \)).
    \end{itemize}
    \item \textbf{Operation timing attributes} -- these values are used with T1/T2 for thermal relaxation errors:
    \begin{itemize}
        \item \textbf{Readout length (ns)} -- measurement duration;
        \item \textbf{Single-qubit gate length (ns)} -- duration of a single-qubit gate;
        \item \textbf{Gate length (ns)} -- duration of a two-qubit gate on a qubit pair (in practice, it is direction-dependent).
    \end{itemize}
    \item \textbf{Single-qubit depolarizing error attributes} are used for depolarizing errors:
    \begin{itemize}
        \item \textbf{ID error}
        \item \textbf{Z-axis rotation (rz) error}
        \item \textbf{\( \sqrt{x} \) (sx) error}
        \item \textbf{Pauli-X error}
        \item \textbf{RX error}
    \end{itemize}
    \item \textbf{Two-qubit depolarizing error attributes} are used for depolarizing errors on qubit pairs (also direction-dependent):
    \begin{itemize}
        \item \textbf{ECR error}
        \item \textbf{CZ error}
        \item \textbf{RZZ error}
    \end{itemize}
\end{itemize}

Although only two QPUs (\textit{ibm\_brisbane} and \textit{ibm\_sherbrooke}) were available for free plan users while conducting practical tests for this paper, it was (and still is) possible to download QPU data from all active devices at no additional cost. With this in mind, if a QPU can be accurately simulated based on its calibration data, any such device can be replicated locally on classical computers free of charge, though existing restrictions of quantum computer simulation should be kept in mind, for example, extensive RAM usage. The approach may be adaptable to non-\textit{IBM} QPUs with sufficiently similar exposed calibration and target information; evaluating that portability is left for future work. 

\subsection{Noise Model Construction from Calibration Data}
\label{noise_model_construction}

Since the calibration data originates from the \textit{IBM Quantum} platform, we use \textit{IBM Qiskit} and \textit{Qiskit Aer} to construct the simulators and noise models through functionality provided by specific classes and functions.

In our pipeline, \textbf{\textit{NoiseModel}} stores created errors in a single object, \textbf{\textit{QuantumError}} describes gate-related CPTP errors, and \textbf{\textit{ReadoutError}} describes classical readout errors. To instantiate \textbf{\textit{QuantumError}} objects, we use the helper functions \textbf{\textit{depolarizing\_error()}} and \textbf{\textit{thermal\_relaxation\_error()}}. We instantiate \textbf{\textit{ReadoutError}} directly from the measurement-assignment probabilities. We use the calibration fields defined in Section 3.4 and insert the resulting errors into the \textit{NoiseModel} object with \textbf{\textit{add\_quantum\_error()}} and \textbf{\textit{add\_readout\_error()}} for the corresponding qubits and qubit pairs.

The choice of these specific error classes and helper functions is supported by the official \textit{Qiskit Aer} documentation, which describes how backend-derived noise is constructed in similar situations. Additionally, the \textit{NoiseModel} class has a method called \textbf{\textit{from\_backend()}} that is used to obtain automatically created noise models from backend calibration properties. Its documentation also mentions the use of the previously described helper functions and classes.

Prior to creating any errors, we pass a list of QPU-specific basis gates as an argument to the \textit{NoiseModel} object, because default settings may add unnecessary gates that cannot be removed afterwards.

\subsection{Coupling Map Reconstruction from Calibration Tables}
\label{coupling_map}

Each QPU also has its own specific coupling map that defines how qubits are connected. Furthermore, \textit{IBM’s} QPUs have directional coupling maps - not all directions between connected pairs are valid. This property is also essential for precise simulation, however, neither the neighboring qubits nor the coupling direction are directly mentioned anywhere in the calibration data. Though there is a workaround - there are usually multiple values stored in every qubit pair-related attribute column, and each of these values has a target qubit mentioned. Thus it is possible to obtain the coupled qubit pairs by combining the current qubit (the control qubit) and the target qubits from these columns -- though it takes additional effort. 

Once the coupling topology is obtained, it can be defined with the \textbf{\textit{CouplingMap}} class as a directed graph of permitted couplings. Since \textit{NoiseModel} objects do not contain coupling maps, they are added to a simulator instance along with the created noise model (e.g., via $AerSimulator(..., noise\_model=..., coupling\_map=...)$).

\subsection{Backend- and Snapshot-Based Noise Models}

In addition to the possibility of creating custom noise models, \textit{Qiskit} also provides functionality for obtaining automatically created noise models and simulators. Both \textit{AerSimulator} and \textit{NoiseModel} classes have the method \textit{from\_backend()} that configures the respective objects to mimic a target backend, similarly to our previously described pipeline in Sections \ref{noise_model_construction} and \ref{coupling_map}.

The backend-derived construction depends on the backend availability under the current IBM Quantum access plan of the user. As an alternative, the module \textit{qiskit\_ibm\_runtime.fake\_provider} provides “fake backends” that mimic IBM Quantum systems using system snapshots. For those, the calibration data will not be up-to-date with the currently available QPUs on the platform. Also, the snapshot coverage can vary across releases/devices (e.g., when the practical tests for this research were conducted, only 7 out of 12 QPUs were available as fake backends).

\subsection{Simulator Configuration Choices}

Without going in depth on the differences of each method, a simple test was conducted to see if it is even worth testing out every valid simulation method separately. The three methods that complied with the previously stated requirements (\textit{statevector}, \textit{density\_matrix}, \textit{matrix\_product\_state}) were used to run all three circuits with 100000 shots, and their result similarities were compared. In all cases the similarity between results from each simulation method was around 97.6\%, which also crosses the previously defined 95\% similarity threshold. Thus we can conclude that, at least in the terms of results, there are no noticeable differences between the simulation methods.

As for the choice of \textit{density\_matrix} --- it was the fastest of the three tested simulation methods.

\subsection{Evaluation Protocol}

This subsection describes the protocol used to evaluate the agreement between QPU executions and the corresponding digital-twin executions. The evaluation is performed at the level of measured state-count outputs. The protocol specifies the similarity metric, the benchmark circuits, the transpilation settings, the shot-budget choice, and the practical data-collection setup used in the experiments.

\subsubsection{Similarity Metric}
\label{similarity_metric}
Since algorithms are typically executed multiple times on quantum computers, which is specified by the number of shots, the results for circuit execution jobs consist of specific state measurement counts, for example: 

$$
[0:531; 1:469]
$$

In this example, ‘0’ and ‘1’ represent measured quantum states, whereas ‘531’ and ‘469’ are specific state count values -- the numbers of times each of the states was measured as the result. 

In order to obtain a specific percentage value that defines the similarity of such results, a custom function was created based on the \textbf{\textit{Weighted Jaccard Similarity}} formula: \cite{jaccard_similarity}

$$
J_w(x, y) = \frac{\sum_{i=1}^n \min (x_i,y_i)} {\sum_{i=1}^n \max (x_i,y_i)}
$$

In simple terms, for each pair of comparable results, the function goes through all measured states and calculates the sums of \textit{min} and \textit{max} state counts from both results. After dividing these two sums accordingly, based on the previously presented \textit{Weighted Jaccard Similarity} formula, and multiplying the result by `100', the required percentage values are obtained.  The function would then return a similarity matrix that consists of the calculated result similarity percentage values. Below is an example of a returned similarity matrix: 

$$
\begin{bmatrix}
100.0 & 67.306.. \\
67.306.. & 100.0
\end{bmatrix}
$$

This approach makes it very easy to read the similarity of the tested results, as Weighted Jaccard provides a normalized and easily interpretable measure of overlap between two count distributions. By directly comparing the frequencies of shared outcomes, it gives a practical similarity score for simulator–hardware validation. As shown in the example, all diagonal values of the matrix are 100, which corresponds to each result being identical to itself. Each row of the matrix shows how one result instance (the one with the value `100') compares to all other result instances. Another characteristic of these similarity matrices is that values around the diagonal line will be mirrored --- they will be the same on both sides of the diagonal line.

\subsubsection{Benchmark Circuits and Transpilation Settings}
To obtain the required results, which could then be compared with the previously described method, there must be a quantum circuit that both the simulator and the real QPU will run. 

\begin{figure}[htp]
    \centering
    \includegraphics[width=12cm]{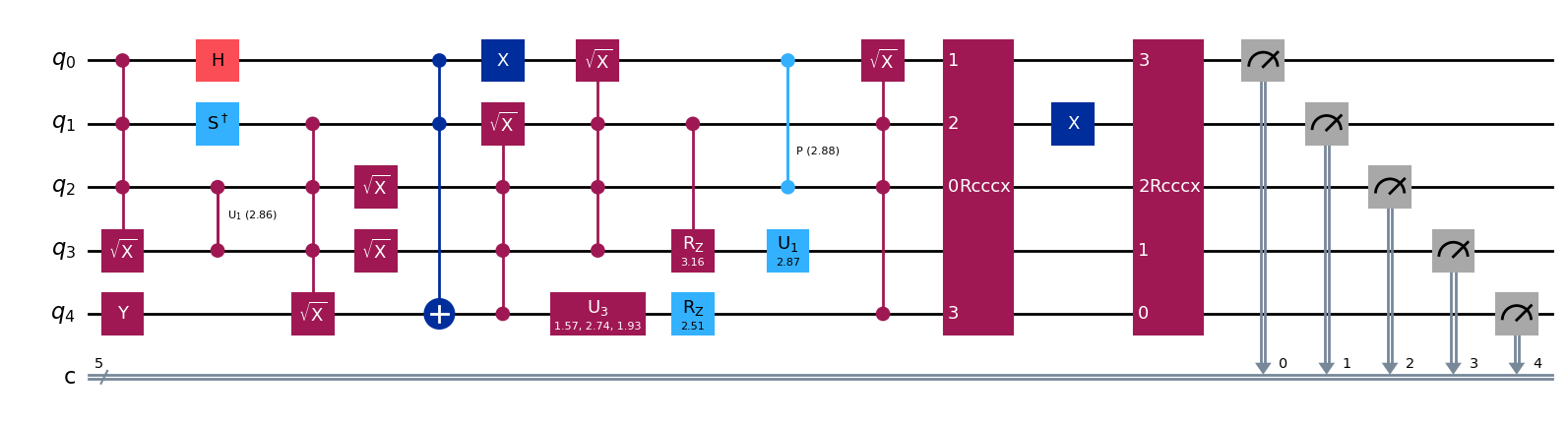}
    \caption{Generated five-qubit quantum circuit with depth 10}
    \label{fig:5q10}
\end{figure}

We decided to use three different randomly generated five-qubit quantum circuits --- created by the \textit{Qiskit} function \textit{random\_circuit()}, 
with circuit depths 10, 20, and 30. As the number of gates increases, the circuit becomes more susceptible to errors, which in turn allows the evaluation of a wider spectrum of different scenarios and their potential impact on the final results. Five-qubit quantum circuit with depth 10 that was used during the conducted practical tests is shown on Fig.~\ref{fig:5q10}. 

Since the circuits are randomly generated, so are their expected results. We performed a preliminary check comparing a noiseless simulation with a noisy simulation from a QPU with the \textit{from\_backend()} method to confirm that the chosen circuits are noise-sensitive. For example, the result difference for the same five-qubit circuit with depth 10 was 43,6\%, confirming that this circuit is valid for further tests. 


Validation practices in quantum computing often rely on characterization and benchmarks based on randomized circuits. This includes randomized benchmarking and randomized circuits of Quantum Volume model \cite{ref_magesan_rb,ref_cross_qv}. We avoided algorithmic benchmarks like Shor-type circuits because these typically have more qubits and require longer execution times. For example, \textit{Shor’s} algorithm from a \textit{Qiskit} tutorial \cite{shors_algorithm} was initially tested, and attempts to execute on a real QPU ended with errors because the process took too long.



Before a quantum circuit can be run on any QPU or specific simulator instance, it must be transpiled accordingly. In simple terms, transpilation ensures required adjustments to the circuit: 
\begin{itemize}
    \item Mapping circuit qubits onto the coupling map of the QPU or simulator; 
    \item Routing the circuit so that it is compatible with the coupling map of the QPU or simulator; 
    \item Translating circuit gates to the basis gates that the QPU or simulator accepts. 
\end{itemize}

When transpiling a circuit, it is also possible to optimize the process, which may help reduce the effects of noise and provide more precise results. However, it is important to note that these improvements are not always guaranteed and noticeable. Optimization can be configured to one of the four available levels, ranging from `0' to `3', which determines the type and extent of additional work done as part of the optimization process. 

Since this would not deviate from the main goal of this research, but instead add additional potentially useful information in the form of test results, it was decided to also test whether different optimization levels would affect the similarity between simulator and QPU results. This meant that each of the three quantum circuits were tested with all four optimization levels.

\subsubsection{Shot Budget Selection}
\label{shot_budget_selection}

The attribute \textit{shots} determines how many times a quantum circuit is executed, and statistical variability of circuit execution results decreases with increasing shot counts. Although higher shot counts typically yield more stable output distributions, it also leads to longer execution time, since quantum processing time scales with the number of shots. During preliminary research, a hard-coded limit of 100000 shots was discovered for both \textit{ibm\_brisbane} and \textit{ibm\_sherbrooke}, which can be obtained with \textit{backend.configuration().max\_shots}.


Even with this information, additional tests were conducted before picking a specific number to ensure the most optimal result similarity as there was no guarantee that 100000 would be either enough or too excessive. 

To answer this, all three randomly generated quantum circuits were each executed 100 times with an automatically generated noise model from a QPU and a specific shot count. The tests began with 1000 shots, then increased to 5000, after which all subsequent iterations would add 5000 more shots until reaching the previously mentioned hard-coded limit. During each iteration, all 100 obtained results were compared with one another, and the highest and lowest similarity values were recorded. 

Table~1 contains some of the recorded lowest and highest result similarities for all three circuits with the optimization level set to `0' (full table is available at Github).

\begin{longtable}{|m{1.5cm}|p{1.5cm}|p{1.5cm}|p{1.5cm}|p{1.5cm}|p{1.5cm}|p{1.5cm}|}

\caption{Obtained min and max similarity values from 100 circuit executions with specific shot counts} \\

\hline
\multirow{2}{*}{\textbf{Shots}} &
\multicolumn{2}{c|}{\textbf{Circuit Depth 10}} &
\multicolumn{2}{c|}{\textbf{Circuit Depth 20}} &
\multicolumn{2}{c|}{\textbf{Circuit Depth 30}}
\\ \cline{2-7}
&
\textbf{Min \%} & 
\textbf{Max \%} & 
\textbf{Min \%} &
\textbf{Max \%} &
\textbf{Min \%} &
\textbf{Max \%}
\\ \hline
\endfirsthead

\hline
\multirow{2}{*}{\textbf{Shots}} &
\multicolumn{2}{c|}{\textbf{Circuit Depth 10}} &
\multicolumn{2}{c|}{\textbf{Circuit Depth 20}} &
\multicolumn{2}{c|}{\textbf{Circuit Depth 30}}
\\ \cline{2-7}
&
\textbf{Min \%} & 
\textbf{Max \%} & 
\textbf{Min \%} &
\textbf{Max \%} &
\textbf{Min \%} &
\textbf{Max \%}
\\ \hline
\endhead
\label{min_max_similarity_values}

1000 &
72.265\% & 
89.214\% & 
73.611\% &
89.214\% &
75.131\% &
89.573\%
\\ \hline
5000 &
87.899\% & 
95.274\% & 
87.160\% &
95.007\% &
87.300\% &
95.160\%
\\ \hline
10000 &
91.296\% & 
96.328\% & 
91.314\% &
96.444\% &
90.949\% &
96.714\%
\\ \hline
15000 &
92.604\% & 
97.109\% & 
92.628\% &
97.381\% &
92.270\% &
96.992\%
\\ \hline
30000 &
94.723\% & 
98.072\% & 
94.514\% &
98.202\% &
94.779\% &
98.104\%
\\ \hline
60000 &
96.299\% & 
98.603\% & 
96.286\% &
98.606\% &
96.174\% &
98.613\%
\\ \hline
100000&
97.022\% & 
98.828\% & 
96.984\% &
98.975\% &
96.817\% &
99.026\%
\\ \hline

\end{longtable}

Based on the results, the most significant improvements occurred between 1000 and 15000 shots. Even though the rate of improvement decreased significantly at higher shot counts, improvement for both highest and lowest similarity values was observable. Additionally, executing one circuit on a QPU with the maximum number of shots took approximately 40 seconds, which was considered an acceptable execution time. Since the hard-coded limit of 100000 shots continued to yield improvements in terms of obtained result similarity, along with the fact that the execution time per circuit was acceptable, there was no reason for not selecting it for further tests.

\subsubsection{Hardware, Data Collection, and Experimental Matrix}

The experiments were conducted on two IBM Quantum QPUs, that were available at the moment of conducting practical tests -- \textit{ibm\_brisbane} and \textit{ibm\_sherbrooke}. For each device, three benchmark circuits were executed at four optimization levels and compared with four simulator-based twin variants, yielding the full experimental matrix used for validation. In total, we performed 24 experiments, each having one hardware job. For each hardware job there are four simulator runs whose outcomes are compared between each other and the related hardware job (this makes 10 comparisons per experiment). As a result, we have 24 hardware jobs and 96 simulator runs.



During data collection, hardware jobs, saved noise data, and measurement counts were organized to reduce the effect of calibration drift over time. With the fact that it was only possible to submit three jobs to the queue simultaneously, along with potential waiting times spanning multiple hours, the jobs had to be divided accordingly so that there would be minimal impact of changing calibration data over time, which would otherwise lead to increased result differences. Thus, it was decided that for every set of three jobs, all three circuits would be submitted to run on a specific optimization level on one of the two devices. 

Along with the instances of noise data, result counts of each job were saved so that they could be later compared with simulator results.



\section{Results and Discussion}
\subsection{Example Similarity Matrix and Interpretation}

The Table~\ref{test_similarity_matrix_example} shows a similarity matrix for one of the 24 tests as an example. The QPU results were compared with simulator results, each with one of the four selected noise data sources. In addition to this, simulator results could also be compared one to another.

\begin{table}[htbp]
\caption{
    Similarity matrix example for 1 of 24 experiments --- for \textit{ibm\_brisbane}, five-qubit circuit with depth 10, and optimization level 3
    \label{test_similarity_matrix_example}
}
\centering
\begin{tabular}{|>{\centering\arraybackslash}m{1.8cm}|>{\centering\arraybackslash}m{1.8cm}|>{\centering\arraybackslash}m{1.8cm}|>{\centering\arraybackslash}m{1.8cm}|>{\centering\arraybackslash}m{1.8cm}|>{\centering\arraybackslash}m{1.8cm}|}

\hline
& 
\textbf{QPU} & 
\textbf{Simulator Instance From QPU} & 
\textbf{Noise Model Instance From QPU} & 
\textbf{\textit{Fake Brisbane}} & 
\textbf{Noise Model From CSV}
\\ \hline
\textbf{QPU} &
100\% &
75.940\% &
76.119\% &
70.136\% &
74.725\%
\\ \hline
\textbf{Simulator Instance From QPU} &
75.940\% &
100\% &
98.078\% &
78.541\% &
95.114\%
\\ \hline
\textbf{Noise Model Instance From QPU} &
76.119\% &
98.078\%&
100\% &
77.996\% &
95.532\%
\\ \hline
\textbf{\textit{Fake Brisbane}} &
70.136\% &
78.541\% &
77.996\% &
100\% &
75.634\% 
\\ \hline
\textbf{Noise Model From CSV} &
74.725\% &
95.114\% &
95.532\% &
75.634\% &
100\%
\\ \hline

\end{tabular}
\end{table}

\subsection{Overall Agreement Distribution}

Experiments produced in total 96 simulator results to analyze (48 for each QPU). All results were grouped into similarity intervals for further analysis, which can be seen in the Table~\ref{grouped_results}.

\begin{table}[t]
\centering
\small
\setlength{\tabcolsep}{4pt}
\caption{All 96 simulator results grouped into similarity intervals.}
\label{grouped_results}
\begin{tabular}{|l|l|l|l|}
\hline
\makecell[l]{\textbf{Similarity} \\ \textbf{Intervals}} &
\textbf{Result count} &
\makecell[l]{\textbf{Percentage of} \\ \textbf{Total Results}} &
\makecell[l]{\textbf{Cumulative} \\ \textbf{Percentage}} \\ \hline

\makecell[l]{Above 95\%} &
\makecell[l]{Total: 7 \\ \textit{ibm\_brisbane}: 7 \\ \textit{ibm\_sherbrooke}: 0} &
\makecell[l]{Total: 7.29\% \\ \textit{ibm\_bris}: 14.58\% \\ \textit{ibm\_sher}: 0\%} &
\makecell[l]{Total: 7.29\% \\ \textit{ibm\_bris}: 14.58\% \\ \textit{ibm\_sher}: 0\%}
\\ \hline

\makecell[l]{Between 95\% \\ and 90\%} &
\makecell[l]{Total: 17 \\ \textit{ibm\_brisbane}: 7 \\ \textit{ibm\_sherbrooke}: 10} &
\makecell[l]{Total: 17.71\% \\ \textit{ibm\_bris}: 14.58\% \\ \textit{ibm\_sher}: 20.83\%} &
\makecell[l]{Total: 25.00\% \\ \textit{ibm\_bris}: 29.17\% \\ \textit{ibm\_sher}: 20.83\%}
\\ \hline

\makecell[l]{Between 90\% \\ and 85\%} &
\makecell[l]{Total: 25 \\ \textit{ibm\_brisbane}: 13 \\ \textit{ibm\_sherbrooke}: 12} &
\makecell[l]{Total: 26.04\% \\ \textit{ibm\_bris}: 27.08\% \\ \textit{ibm\_sher}: 25.00\%} &
\makecell[l]{Total: 51.04\% \\ \textit{ibm\_bris}: 56.25\% \\ \textit{ibm\_sher}: 45.83\%}
\\ \hline

\makecell[l]{Below 85\%} &
\makecell[l]{Total: 47 \\ \textit{ibm\_brisbane}: 21 \\ \textit{ibm\_sherbrooke}: 26} &
\makecell[l]{Total: 48.96\% \\ \textit{ibm\_bris}: 43.75\% \\ \textit{ibm\_sher}: 54.17\%} &
\makecell[l]{Total: 100\% \\ \textit{ibm\_bris}: 100\% \\ \textit{ibm\_sher}: 100\%}
\\ \hline
\end{tabular}
\end{table}

Based on the information presented in Table~\ref{grouped_results}, the following observations can be made:
\begin{itemize}
    \item Approximately half of all simulator results (51.04\%) were with a similarity of 85\% or higher, when compared to respective QPU results;
    \item Out of 96 simulator results, 7 exceeded the highest result similarity threshold of 95\%. All 7 results were for the \textit{IBM Quantum} QPU \textit{ibm\_brisbane}; 
    \item If compared, there are more results with higher similarity for \textit{ibm\_brisbane} than there are for \textit{ibm\_sherbrooke}. 
\end{itemize}

\subsection{Which Twin Variant Matches Best}

Table~\ref{grouped_tests_by_sources} highlights how often a specific noise data source produced the highest similarity value across all four tested sources, when compared to the QPU results.

\begin{table}[htbp]
\caption{
    All 24 tests grouped by specific noise data sources with highest result similarity to a QPU.
    \label{grouped_tests_by_sources}
}
\centering
\begin{tabular}{|p{4cm}|p{3cm}|p{4cm}|}

\hline
\textbf{Noise Data Source} &
\textbf{Times With Highest Similarity} &
\textbf{Times Grouped by QPU}
\\ \hline
CSV calibration data &
13 / 24 &
\makecell[l]{\textit{ibm\_brisbane}: 6 \\
\textit{ibm\_sherbrooke}: 7}
\\ \hline
Noise model or simulator instance from backend (QPU) &
7 / 24 &
\makecell[l]{\textit{ibm\_brisbane}: 5 \\
\textit{ibm\_sherbrooke}: 2}
\\ \hline
Fake backends (FakeBrisbane and FakeSherbrooke) &
4 / 24 &
\makecell[l]{\textit{ibm\_brisbane}: 1 \\
\textit{ibm\_sherbrooke}: 3}
\\ \hline

\end{tabular}
\end{table}

\subsection{Sensitivity to Optimization Level and Device}

There were some interesting recurring observations that are worth mentioning: 
\begin{itemize}
    \item When compared to other simulator results, instances that used fake backends (\textit{FakeBrisbane} and \textit{FakeSherbrooke}) as noise sources had noticeably lower similarity ratings if the optimization level was above 0; 
    \item In general, optimization level changes affected the resulting similarity values in varying ways, making it difficult to draw definitive conclusions. The following examples illustrate this behavior: 
    \begin{itemize}
        \item Optimization level changes from 0 to 1 lead to increased similarity values. Though, at the same time, the similarity values decreased with level 2 and 3 (\textit{ibm\_brisbane} QPU, five-qubit circuit with depth 10); 
    \item Optimization level changes from 0 to 1 lead to decreased similarity values. (\textit{ibm\_sherbrooke} QPU, same circuit as previous example). 
\end{itemize}
\item Backend-derived simulator and noise-model twins produced almost identical results, with the similarity consistently being around 95\%. Furthermore, when comparing the two variant results to the QPU, both similarity values differed by only about 1\%;
\item In cases with the \textit{ibm\_brisbane} QPU, while comparing results from CSV calibration data-based twins to backend-derived simulator and noise-model variants, the similarity almost always exceeded 95\%. In the very few cases where this did not occur, the similarity was still very close to 95\%. However, this cannot be said about the \textit{ibm\_sherbrooke} QPU, as the similarity of the same twin variant results neither reached 95\% nor was close enough to it;
\item The highest result similarity values were always encountered with either optimization level 0 or 1. Only one exception occurred - running the five-qubit quantum circuit with depth 30 at optimization level 2. 
\end{itemize}


\subsection{Implications for Calibration-Based Twins}

The results indicate that CSV-based twins are a practical and often strong-performing option despite their higher implementation effort, while backend-derived twins remain useful baselines.

Twin similarity should be validated per device, since performance on one QPU does not guarantee comparable results on another. Moreover, results vary with transpilation settings, and so the transpiler configuration should be treated as part of the digital-twin definition and validation setup.

In several test conditions, simulator outputs from different twin variants are highly similar to each other while remaining noticeably less similar to the corresponding QPU output. This implies that high inter-simulator similarity is not a substitute for QPU validation.

The reported threshold outcomes should be interpreted alongside the full similarity distribution. For practical use, the required similarity level may depend on the intended application.

\subsection{Threats to Validity and Limitations}
The study used the platform maximum of 100000 shots. Higher shot counts might further improve agreement, but they were outside the practical limits of the experimental setup.

Even though it was not tested in the scope of this paper, circuits with different qubit counts could potentially give different results.

Along with this, the transpilation process may potentially play a role in determining the similarity. As was brought up in Section 3.9, optimization affects the types of changes done to a circuit before it finishes the transpilation process, and some changes might not be consistent when transpiling the same circuit multiple times with a higher optimization level (2 or 3). Because transpilation for the QPUs and simulators was done independently, there is a possibility that there were some differences that might have affected the final results.   

Assumed noise model class has limitations, and more complex (possibly non-Markovian) effects can affect the difference between the outcomes of simulators and hardware \cite{ref_onorati_noisefit}. Models based only on calibration can be improved by fitting parameterized models from circuit-execution data, potentially improving distribution-level agreement \cite{ref_ji_noiseml}.


\subsection{Conclusion}

The results of this study show that calibration-based digital twins of IBM Quantum hardware are feasible under the tested validation protocol. Twins constructed from downloadable calibration CSV data were often the strongest performers, while backend-derived twins provided competitive and practical baselines. The findings also show that agreement depends on the target device and on the transpilation settings, so validation should be carried out for the intended execution setup rather than being assumed to transfer automatically across devices. The conclusions are limited to the tested devices, circuits, shot budget, and noise-model assumptions. Future research should extend the analysis to broader circuit classes, different qubit counts, and richer calibration-aware noise models.


\section*{Acknowledgment}
The work was supported by Latvian Quantum Initiative under European Union Recovery and
Resilience Facility project no. 2.3.1.1.i.0/1/22/I/CFLA/001.


\begin{thebibliography}{8}



















\bibitem{ref_nguyen_qcloud}
Nguyen, H.T. et al.: Quantum Cloud Computing: A Review, Open Problems, and Future Directions. arXiv preprint arXiv:2404.11420 (2024). \doi{10.48550/arXiv.2404.11420}

\bibitem{ref_romero_noisevalidation}
Romero-\'Alvarez, J. et al.: A Noise Validation for Quantum Circuit Scheduling Through a Service-Oriented Architecture. In: International Journal of Software Engineering and Knowledge Engineering \textbf{34}(09), pp. 1371--1386 (2024). \doi{10.1142/S0218194024410018}

\bibitem{ref_cicero_simulation}
Cicero, A. et al.: Simulation of Quantum Computers: Review and Acceleration Opportunities. In: ACM Transactions on Quantum Computing \textbf{7}(1), pp. 1--35 (2025). \doi{10.1145/3762672}

\bibitem{ref_perez_ibmreliability}
P\'erez Ant\'on, R. et al.: Reliability of IBM's Public Quantum Computers. In: International Journal of Interactive Multimedia and Artificial Intelligence \textbf{9}(3), pp. 155--163 (2025). \doi{10.9781/ijimai.2023.04.005}

\bibitem{ref_mueller_digitaltwin}
M{\"u}ller, R. et al.: Towards a Digital Twin of Noisy Quantum Computers: Calibration-Driven Emulation of Transmon Qubits. arXiv preprint arXiv:2504.08313 (2025). \doi{10.48550/arXiv.2504.08313}

\bibitem{ref_ibm_noisemodels}
IBM: Build Noise Models. IBM Quantum Documentation. Available at: \url{https://quantum.cloud.ibm.com/docs/en/guides/build-noise-models},  last accessed 2026/03/09

\bibitem{ref_ibm_backenddetails}
IBM: View Backend Details. IBM Quantum Documentation. Available at: \url{https://quantum.cloud.ibm.com/docs/en/guides/qpu-information},  last accessed 2026/03/09

\bibitem{ref_luo_digitaltwinclouds}
Luo, W. et al.: A Digital Twin of Scalable Quantum Clouds. In: Proceedings of the 39th ACM SIGSIM Conference on Principles of Advanced Discrete Simulation, pp. 165--175 (2025). \doi{10.1145/3726301.3732296}

\bibitem{ref_cloudslab_iquantum}
Cloudslab: iQuantum -- A Toolkit for Modelling and Simulation of Quantum Computing Environments. GitHub repository. Available at: \url{https://github.com/Cloudslab/iQuantum}, last accessed 2026/03/11

\bibitem{ref_bertomeu_maestro}
Bertomeu, O. et al.: Maestro: Intelligent Execution for Quantum Circuit Simulation. arXiv preprint arXiv:2512.04216 (2025). \doi{10.48550/arXiv.2512.04216}

\bibitem{jaccard_similarity}
Ultipa: Jaccard Similarity. In: Graph Analytics \& Algorithms, Ultipa Documentation, v5.3. Available at:
\url{https://www.ultipa.com/docs/v5.3/graph-analytics-algorithms/jaccard-similarity}, last accessed 2026/03/13

\bibitem{ref_magesan_rb}
Magesan, E., Gambetta, J.M., Emerson, J.: Scalable and Robust Randomized Benchmarking of Quantum Processes. In: Physical Review Letters \textbf{106}, p. 180504 (2011). \doi{10.1103/PhysRevLett.106.180504}

\bibitem{ref_cross_qv}
Cross, A.W. et al.: Validating Quantum Computers Using Randomized Model Circuits. In: Physical Review A \textbf{100}, p. 032328 (2019). \doi{10.1103/PhysRevA.100.032328}

\bibitem{shors_algorithm}
IBM: Shor’s algorithm. IBM Quantum Documentation. Available at: \url{https://quantum.cloud.ibm.com/docs/en/tutorials/shors-algorithm}, last accessed 2026/03/07


\bibitem{ref_onorati_noisefit}
Onorati, E., Kohler, T., Cubitt, T.S.: Fitting Quantum Noise Models to Tomography Data. In: Quantum \textbf{7}, p. 1197 (2023). \doi{10.22331/q-2023-12-05-1197}

\bibitem{ref_ji_noiseml}
Ji, Y. et al.: Data-Efficient Quantum Noise Modeling via Machine Learning. arXiv preprint arXiv:2509.12933 (2025). \doi{10.48550/arXiv.2509.12933}

\end{thebibliography}
\end{document}